# Spurious Frequencies in the Kepler Short Cadence Data

A. S. Baran

Mt. Suhora Observatory of the Pedagogical University, ul. Podchorążych 2, 30-084 Cracow, Poland
email: andy@astro.as.up.krakow.pl
Department of Physics, Astronomy, and Materials Science, Missouri State University, Springfield, MO 65897, USA
email: abaran@missouristate.edu

ABSTRACT

We present our search for artifacts in the Kepler short cadence data using a commonly known Fourier technique. We analyzed data on a monthly basis searching for a possible correlation between artifacts and the events attributed to the spacecraft as potential sources of the spurious frequencies. We defined a peak to be an artifact if it shows in at least two, yet preferentially most of the stars, during a given month. Besides the commonly known long cadence comb we found a periodic appearance of another two combs, one single artifact and very strange wide artifacts roaming between 10 c/d and 35 c/d. These artifacts evolve on a yearly basis (four of Kepler's rolls) and we may only speculate that their sources are in the reaction wheels since they are the only moving parts or temperature variation. The orientation of the spacecraft is likely excluded from the possible sources. More resources are needed to provide a definite explanation of the artifacts.

## 1. Introduction

The ultimate goal of the Kepler spacecraft is to search for planetary transits in the photometric data of cool stars. It continuously monitors a vast number of stars in a large field of view looking for small dips in the light curves (Borucki *et al.* 2010). Such monitoring is well appreciated by the pulsating stars community for which these data became irreplaceable. Undoubtedly the Kepler spacecraft has contributed to breakthroughs in every type of pulsating stars owing to the unprecedented quality and high duty-cycle data.

To analyze data of pulsating stars, where the variability is strongly periodic, the Fourier technique is used. This includes a calculation of the amplitude spectra and prewhitening to precisely derive the properties of a flux variation. The Fourier technique does not distinguish between a signal intrinsic to the star and the spacecraft. It means that if the spacecraft is influencing our data periodically it will occur as any other signal in the amplitude spectra. Depending on the source of the artifacts the profiles of "fake" frequencies may be different from what is expected for a specific type of pulsator.

Naturally, there are many sources of the "fake" signal on board of the spacecraft. The electronics controlling the scientific instrument and reaction wheels (the only moving parts of the spacecraft) can interfere with the CCDs and contribute to the spurious signal. Since the spacecraft is not thermally stabilized a temperature variation may also contribute to the artifacts.

The goal of our analysis was to find a spurious signal in the amplitude spectra of stars monitored in the short cadences and, if possible, provide the sources of those artifacts. We examined all the quarters of data available to us at the time of the analysis (Q0–14), and listed all suspicious frequencies, repeatable in many stars, we found in the amplitude spectra calculated from monthly chunks. Surprisingly we detected families of artifacts and correlations between their types and months/quarters over which the data were collected. This may provide the sources of artifacts and help to undertake appropriate action for their mitigation. The list of known artifacts will also serve as a handy resource of "fake" signals, helping scientists to avoid misinterpretation of their results based on the Kepler data. In Fig. 1 we show how the artifacts can influence or hide among a real signal.

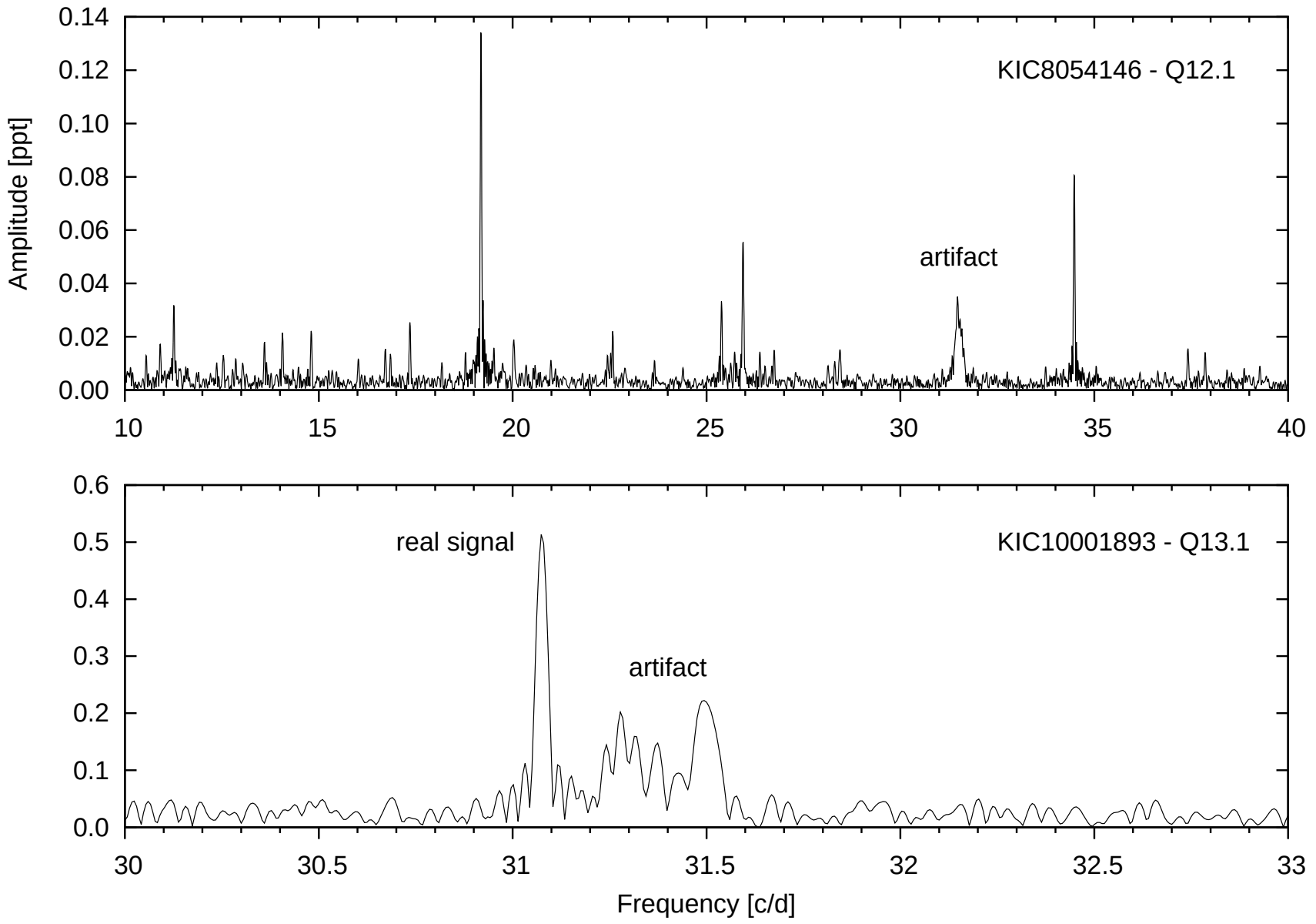

Fig. 1. Close-ups of the amplitude spectra of δ Sct type star KIC 8054146 (*top panel*) and the pulsating subdwarf B star KIC 10001893 (*bottom panel*). They show complex wide artifacts next to the real signal associated with pulsation modes.

## 2. Our Sample and our Strategy

The Kepler spacecraft collects data in two modes, short and long cadences. The short cadence (SC, often called a one-minute cadence) is a pile up of nine exposures of 6.02 s each separated by an overhead of 0.52 s. In total, it gives a short cadence point which is 58.85 s long. The long cadence (LC) is another pile up of 30 SC points leading to 1765.5 s and it is commonly known as a thirty-minute cadence. The cadences easily translate into frequency space giving the Nyquist frequencies for both SC and LC. These frequencies limit our frequency range in which we can detect usable signal for further analysis. In case of LC, the Nyquist frequency equals 24.47 c/d for time series evenly sampled in the spacecraft time. The barycentric correction applied to the spacecraft time shifts cadences by a non-linear, in time, offset producing an uneven time series. This has a strong impact on the Nyquist frequency which does not remain fixed. It varies in frequency and potentially filters the signal originating in the sub- and super-Nyquist regions (Murphy, Shibahashi and Kurtz 2013). Unfortunately, this filtering will likely be confusing in our search for artifacts and we decided not to include the super-Nyquist region in the amplitude spectra calculated from LC data. Instead, we solely focused on SC data extending the workable range of a frequency up to 734.07 c/d. This is also supported by the fact that most of the pulsators have periodicities at frequencies higher than 24.47 c/d, so it is a natural conclusion to account for that high frequency region in our search.

Limited by on-board storage and monthly downlink of data, the total number of stars is about 170 000 monitored in LC, while a maximum of only 512 stars can be observed in SC during a specific month. The number of SC stars observed over one quarter fluctuates since one star may be observed only during one month allowing more targets to be monitored within a quarter. Therefore, our search for artifacts is based on at least 512 stars observed every quarter.

Since the spacecraft is re-oriented for monthly downlink, leading to a disturbance in the data collection process, we separated data into months. During the collection time there is not much impact from the ground control and we suspected that only a movement of the entire spacecraft, including quarterly rolls, or its instability may “inject” a spurious signal into the data. However, if the reaction wheels and the electronics contribute to the fake signal pool it may continuously affect our science data.

We downloaded all SC data available to the public (Q0–14) from Mikulski Archive for Space Telescopes. We used fluxes derived from an optimal aperture and called Simple Aperture Photometry (SAP). These fluxes were $3\sigma$ clipped and detrended using a third order spline fit calculated from 0.5 day bins. Next, we calculated the amplitude spectra on a monthly basis and examined them by eye.

There are two ways to identify artifacts. If we know the source of a disturbance and we can anticipate its time scale then we can associate it with the peaks in the amplitude spectra. The drawback of this option is that we do not quite understand how the mutual interferences on-board the spacecraft work. Moreover, there are real signals in stars which exist very close to the artifacts which means that not all the peaks at those anticipated frequencies can be artifacts.

Instead, we can hunt for the same (in frequency and shape) peaks in the amplitude spectra detected in a large sample of stars. We do not anticipate that the stars are correlated and they should not show the same periodicities. If we find some suspicious periodicity in the amplitude spectra of many stars it will mean that they are not intrinsic to the stars and then we can make an attempt to correlate that periodicity with an event on-board.

## 3. Artifacts Present in Short Cadence Data

The most common artifacts in SC data are those related to the long cadence readout time (LC artifacts). The parent peak of the LC comb (hereafter LC) appears at the long cadence period, that is at 48.94 c/d in frequency. All the other peaks pop up at consecutive harmonics of the parent peak. It is not yet understood but the most significant peaks show up at the $6^{\rm th}$, $7^{\rm th}$ or $8^{\rm th}$ harmonics and not at the parent peak. In fact, the parent peak is typically barely detected, if at all. The LC artifact peaks are usually the strongest in the amplitude spectrum and are well documented (Christiansen *et al.* 2013). These artifacts are present in all months/quarters.

Other known and explained artifacts are located at the frequency of 0.33 c/d and 7.5 c/d and are caused by the reaction wheels and reaction wheels housing temperature, respectively. Because we detrended data, this process has removed any signal below 1 c/d, therefore we did not notice 0.33 c/d artifacts, however we noticed 7.5 c/d in Q1. The amplitude of the latter one was small and we did not detect it in Q3 and onward, as should be the case according to the information given in Kepler Data Release 12 (DRN12) in section 5.12. There are other artifacts which are not explained thus far and are listed in DRN12. The list was prepared based on data up to Q7 and does not provide the details of the specific artifacts. Only a few of them were noticed to have changed their frequencies between months or quarters. To infer more information on an artifact's behavior we undertook a systematic artifact hunt separately for each quarter/month aiming at a possible explanation of their sources. Our aim was not fully obtained, however it helped to consolidate a few separately listed artifacts in DRN12 and provided a month to month variation of one of the wide artifacts likely bringing us closer to its explanation. More resources are needed to fully investigate its source.

In total, we found and discussed five different types of unknown artifacts which were commonly found through many if not all quarters. Some of the artifacts have been already presented by Baran *et al.* (2011). The artifacts can be divided into the following groups: a wide artifact below 20 c/d (hereafter 20-), another wide one between 20 and 35 c/d (hereafter 20+), a comb with a separation of about 35.7 c/d (including that peak) (hereafter U), another comb with a separation of 48.94 c/d (like the LC comb) (hereafter LC') and a single peak denoted as W. In general, the artifacts do not remain at fixed frequencies but float on a monthly basis.

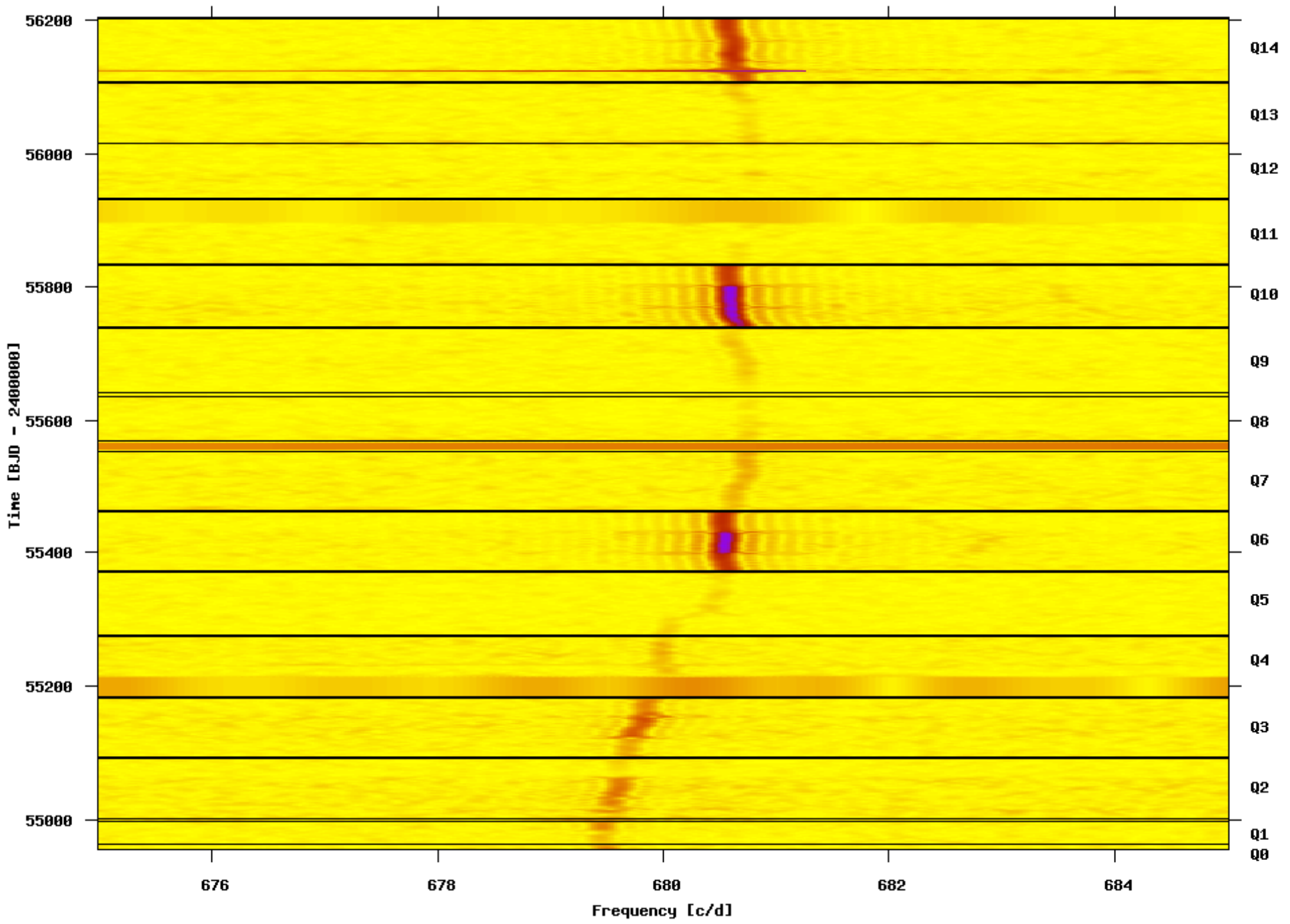

Fig. 2. Time frequency diagram showing the evolution of the U18b artifact around 680 c/d. Horizontal diffused lines are caused by gaps in data while thin black lines separate quarters of data (here and in Fig. 3 and Fig. 20).

The U sequence of artifacts is a conundrum. It shows up almost each month, however its appearance strengthens every four quarters (Q2,6,10,14). Moreover, the separation corresponds to the frequency of a peak which sometimes pops up in the amplitude spectra and most likely we could call it a parent peak (similar to the LC comb). If that peak changes its frequency then the separation in a sequence is adjusted to that frequency. We have also found a correlation between the U artifacts and the shape of the time series data (a description of Q1 data below). Apart from the LC comb, the U sequence shows the strongest peaks close to the Nyquist frequency. For reference, we listed all the frequencies of the U sequence in Table 1. It includes two sequences of artifacts denoted with letters "a" and "b". Since the components of these sequences change their frequencies [c/d] we only provided the values obtained from Q6.2 data. For other quarters they may change by up to 1 c/d. Only those components with superscript d in Table 1 were ever detected. In Fig. 2 we show the strongest component of the b-sequence, U18b. Basically, this diagram is a pile-up of amplitude spectra calculated from 5-day bins which are shifted by one day. The artifact clearly changes its frequency. The change can be periodic on a long ($>4$ years) time scale. Every four quarters the components become very strong. In Fig. 3 another two components, U8a and U8b, are shown. These are smaller in amplitudes and show up only when U18b is very strong. During other quarters they are undetected. They change their frequencies during and between quarters in the opposite way to each other and they look like a mirror view symmetric with respect to a half distance between them. Supposedly, this behavior may suggest that these two components belong to two potentially different, yet somehow related, sequences.

T a b l e 1

List of the U artifacts based on the values detected in Q6.2

| ID | a | b | ID | a | b |
|---|---|---|---|---|---|
| U0 | 35.7[d] | 37.94 | U10 | 392.7 | 394.94[d] |
| U1 | 71.4[d] | 73.64 | U11 | 428.4[d] | 430.64 |
| U2 | 107.1[d] | 109.34 | U12 | 464.1 | 466.34 |
| U3 | 142.8[d] | 145.04 | U13 | 499.8[d] | 502.04[d] |
| U4 | 178.5 | 180.74 | U14 | 535.5[d] | 537.74 |
| U5 | 214.2 | 216.44 | U15 | 571.2 | 573.44[d] |
| U6 | 249.9 | 252.14 | U16 | 606.9[d] | 609.14[d] |
| U7 | 285.6 | 287.84 | U17 | 642.6 | 644.84[d] |
| U8 | 321.3[d] | 323.54[d] | U18 | 678.3 | 680.54[d] |
| U9 | 357.0[d] | 359.24 | U19 | 714.0 | 716.24[d] |

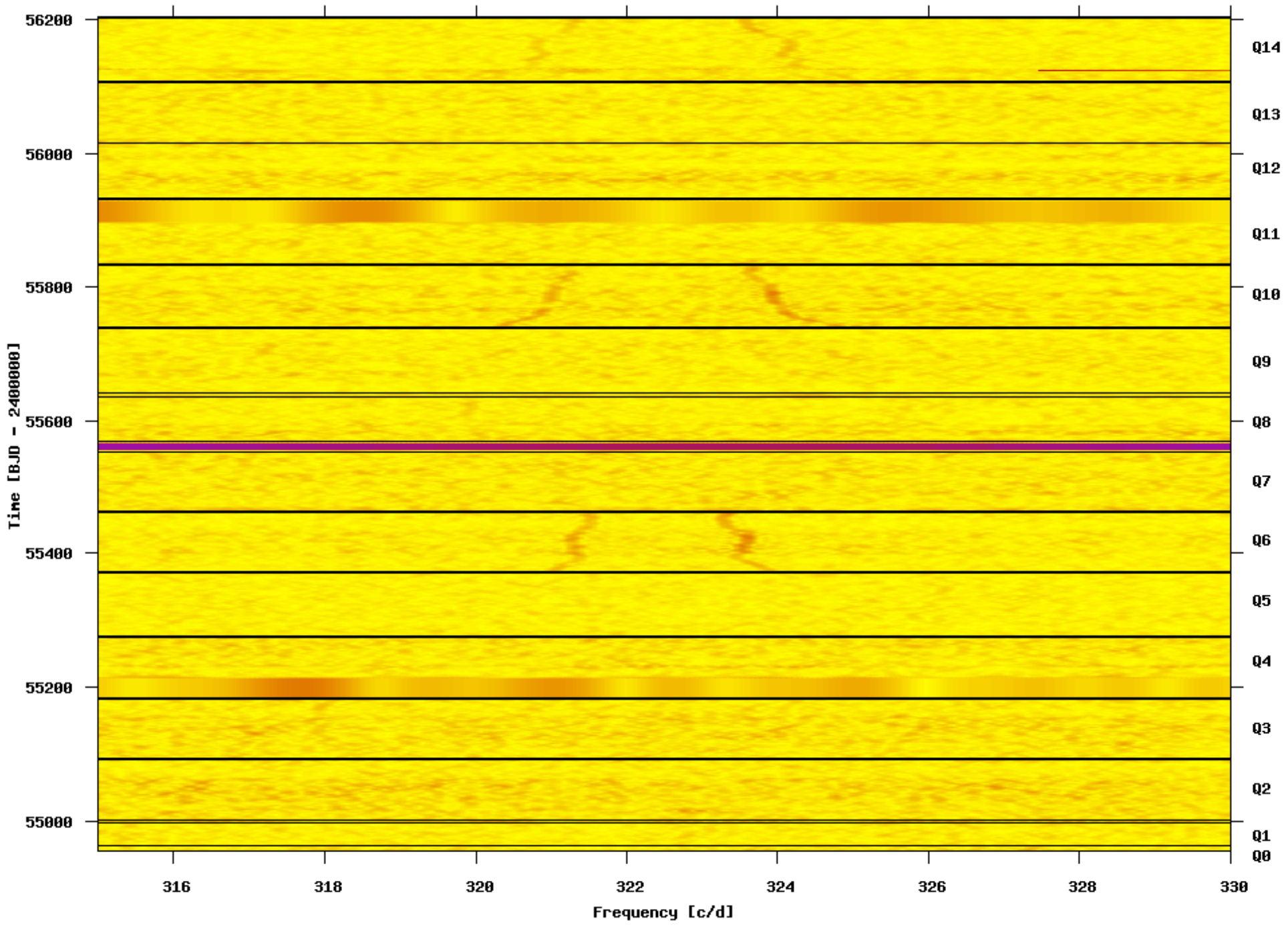

Fig. 3. Time frequency diagram showing the evolution of two artifacts, U8a and U8b, around 320 c/d.

The LC' comb is very unexpected and we have no explanation of this feature besides the fact that the separation is 48.94 c/d so the LC readout likely plays a role. The components of the LC' comb do not appear at the frequencies of the regular LC comb but are shifted by almost 7 c/d. The LC' includes a few peaks only, however the peaks appear in the frequency region where regular LC components are the strongest (400–500 c/d). They change their frequencies by up to 2 c/d. We listed LC' components in Table 2. The sequence is plotted in Fig. 8 where we show artifacts of Q3.

Table 2

List of the LC' components [c/d] for two different months of data to show their frequency change

| ID | Q3.3 | Q7.3 |
|---|---|---|
| $LC'_1$ | 384.12 | 382.61 |
| $LC'_2$ | 433.06 | 431.55 |
| $LC'_3$ | 482.00 | 480.49 |
| $LC'_4$ | – | 529.43 |

A single peak sporadically appearing in some of the stars at around 155–158 c/d was also marked as an artifact (hereafter W). It usually shows up when the U artifacts are strong. Even though it does not seem to belong to the U sequences, origins of U and W artifacts may be the same.

Two wide artifacts identified between 10 c/d and 35 c/d are significantly changing their frequencies. The change is periodic and it is possible that these two artifacts can be the same structure evolving in frequency with time. The shapes of the artifacts are very diversified depending on the month. The shapes may be as wide as 10 c/d and as narrow as a single peak. More description is provided in Section 4.

Below we present a description of the results of our artifact hunt in each quarter of data.

Q0: This was only a 10 day long phase of data collection. We found a wide artifact between 29.2 c/d and 30.7 c/d, as well as sharp peaks U17b-19b. The latter peaks create an equidistant sequence with a separation of 36.4 c/d. This separation is a bit different from the one listed in Table 1. It clearly shows that these artifacts do not remain stable and there must be a process which influences their frequencies. We show those two types of artifacts (the wide one and sharp peaks) in Fig. 4.

Q1: This phase contains one month only. Similarly to Q0 only three different types of artifacts were found, the U sequence with an average separation of 36.4 c/d, the wide artifact between 20.5 c/d and 29 c/d and the W artifact at 158.15 c/d. The first two artifacts are shown in Fig. 5. In Fig. 6 we present close-ups of the light curves of KIC 7022603 and KIC 6106120 which show folded outlines of flux distributions. This shape commonly appears in the light curves of the analyzed stars whenever the U sequence shows up in the amplitude spectra. It may suggest that these two processes are related. A periodicity of that folded outline is around

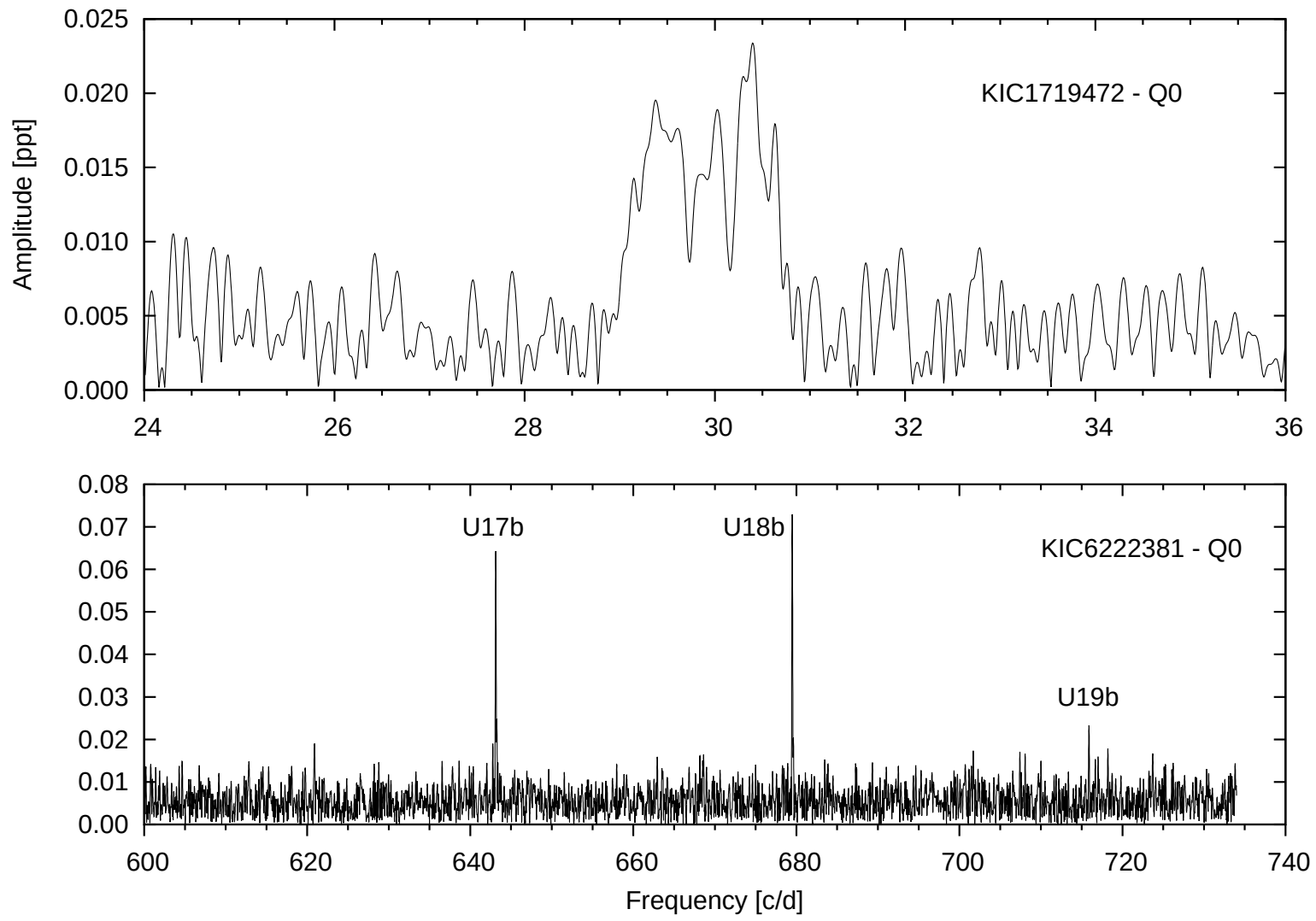

Fig. 4. Close-ups of two amplitude spectra of KIC 1719472 (*top panel*) and KIC 6222381 (*bottom panel*) calculated from Q0. They present two types of artifacts we found, the wide 20+ and a part of the U sequence.

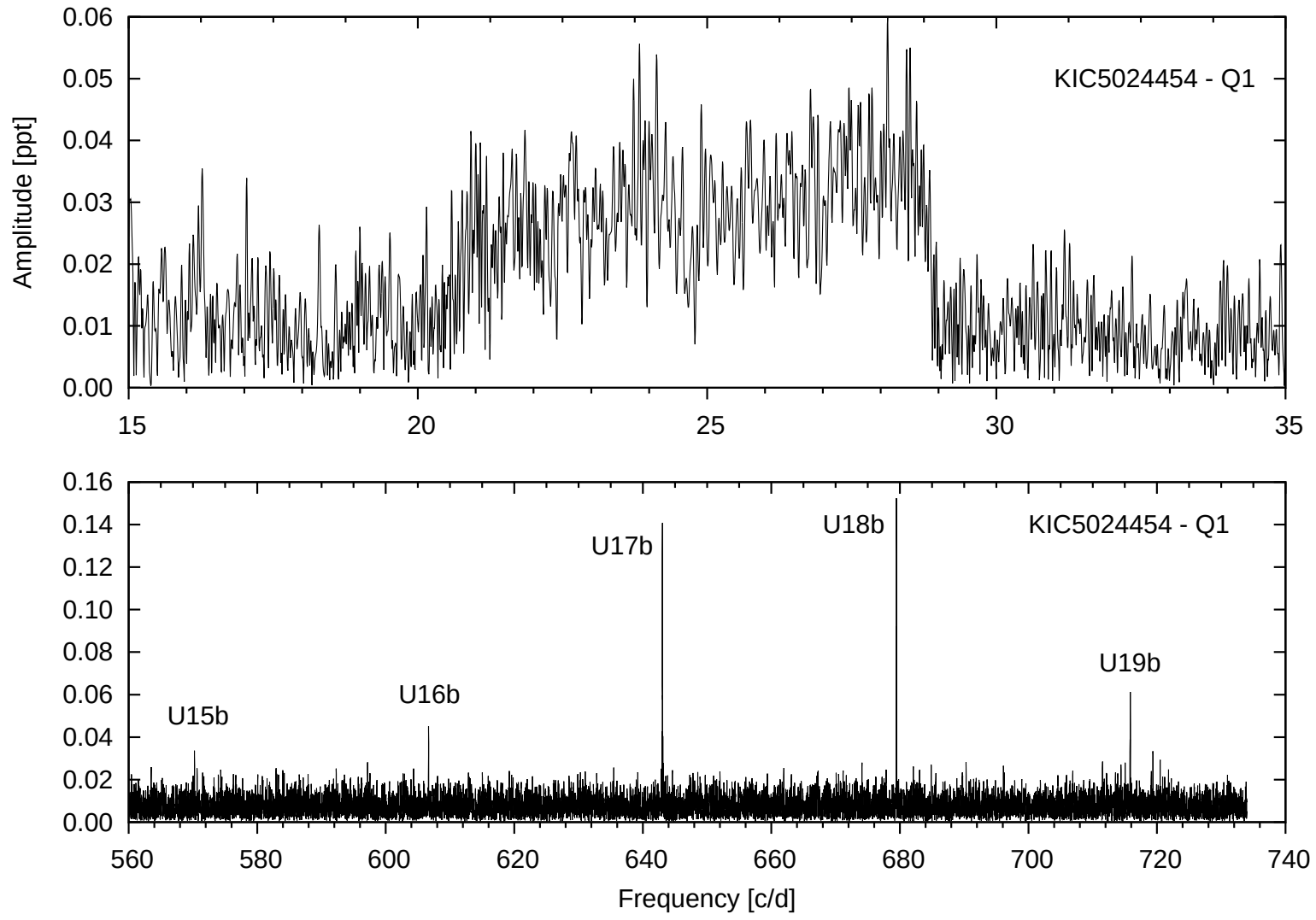

Fig. 5. Close-ups of the amplitude spectrum of KIC 5024454 calculated from Q1 showing the wide 20+ artifact (*top panel*) and the U sequence (*bottom panel*).

3 days. This is close to the timescale of the reaction wheels' desaturation and a temperature variation measured in different parts of the spacecraft (Fig.10 in Christiansen *et al.* 2013). In addition, the outline is always narrower during the process of desaturation. We can only suggest (without any definite conviction) that reaction wheels or temperature variation can be responsible for the U sequence artifacts.

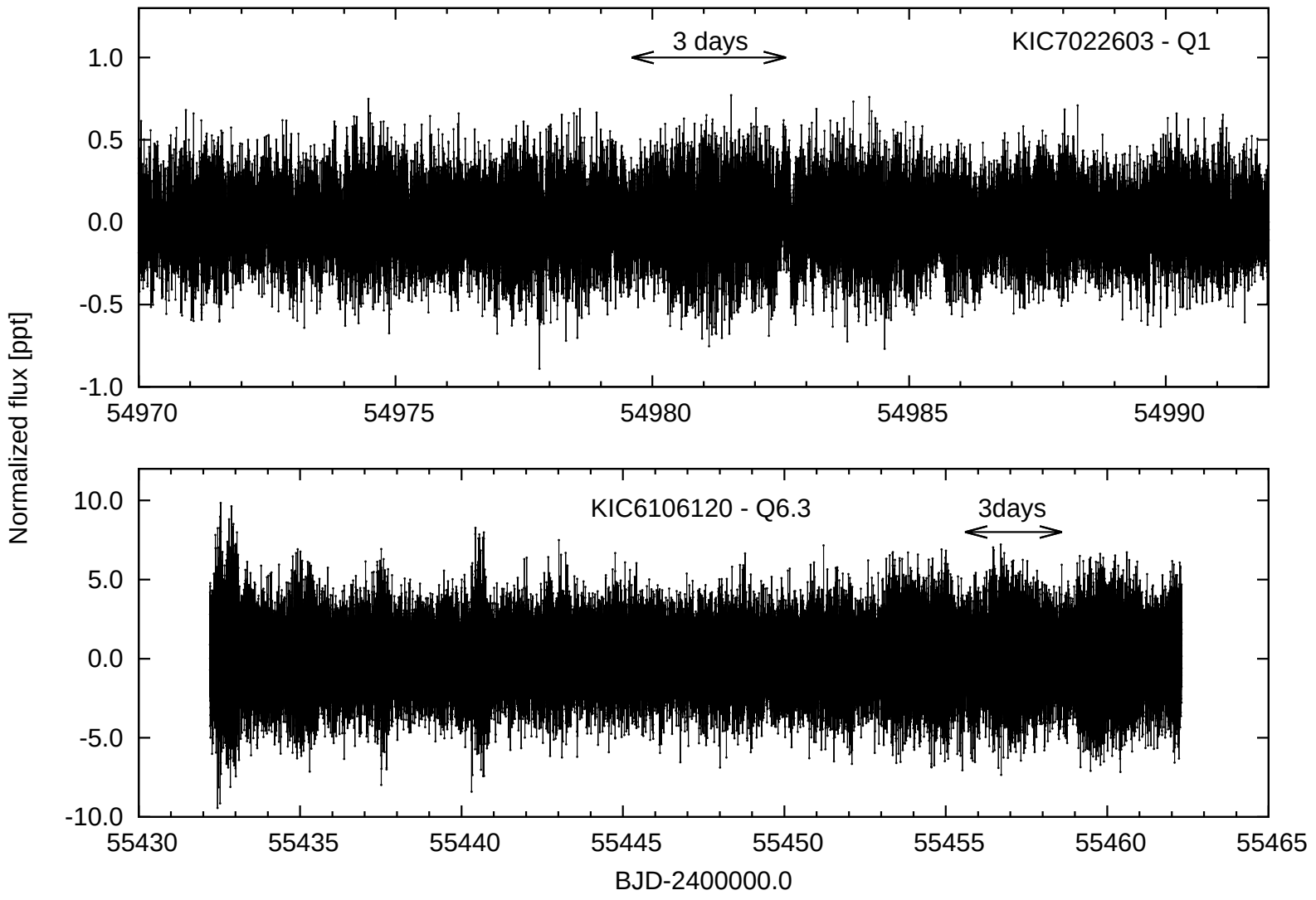

Fig. 6. Close-ups of the light curves of KIC 7022603 obtained during Q1 and KIC 6106120 (Q6.3) showing a folded outline of a flux distribution. The arrows point at the regions where the flux distribution becomes the narrowest which coincidences with reaction wheel desaturation. It repeats every 3 days.

Q2: This was the first full quarter split into three months by data transfer to the Earth. We inspected these three months separately and found that there are three types of artifacts in this quarter. A wide 20-, U sequence and LC' sequence between 300 c/d and 500 c/d. The wide 20- artifact changes between months and contains harmonics. It is very wide in Q2.1 covering almost 6 c/d, while during the next two months it narrowed down to 1 c/d, however it changed its frequency to 17 c/d and then 16 c/d. Harmonics are also noticeable. In Fig. 7 we present the wide artifact and the LC' comb. The U sequence is similar to the ones shown in the previous figures and we do not include them in figures for other quarters unless there are additional features we want to show. During Q2 the U sequence is very intense which suggests that the process causing these artifacts is very strong. The harmonics of wide 20- artifacts are very common features for all quarters and their detection depends on the amplitude of the parent hump. We will not show harmonics for other quarters, however, a reader should be aware of the existence of any peaks which may show up at the multiples of the frequency of the parent peak.

Q3: This quarter is again dominated by the wide artifact 20+, which varies from month to month, and some cases of the U sequence. In addition, we detected the LC' comb. We present all artifacts in Fig. 8. It is worth noting that the wide artifact is continuously changing between months. It is 22–27 c/d during month 1, then 27–31.5 c/d during month 2 and finally sharp but split peaks at 31.5–31.7 c/d. In KIC 4995049 we detected the following artifacts: W at 157.8 c/d and 384.37 c/d, 433.3 c/d, 482.52 c/d and 531.19 c/d, which are the part of the LC' comb.

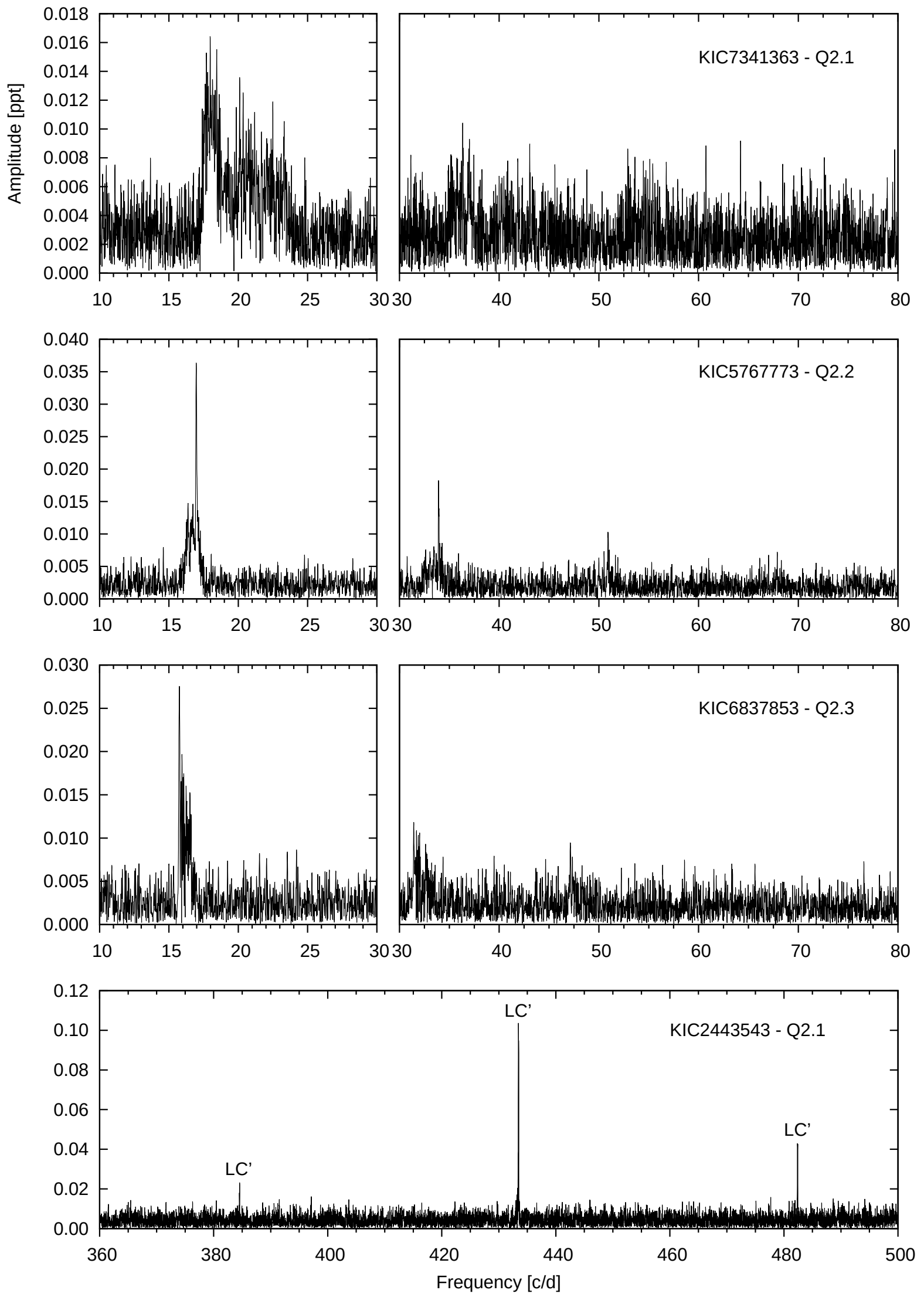

Fig. 7. Close-ups of the amplitude spectra of a few stars calculated from Q2 to show the wide 20-artifact (*top three panels*) and the LC' comb (*bottom panel*).

Q4: In this quarter we found the U sequence, the LC' comb and the wide 20+ artifacts. We show the latter artifact in Fig. 9. The frequency of this artifact is not changing significantly between months, however it differs in shape and remains consistent during a given month. The U sequence is not very common. The artifact W is now at 157.14 c/d.

Q5: Only a few cases of the U sequence were found. Many stars show the wide artifact 20+ which is significantly changing in shape from month to month. Apart from Q3, where that artifact increased its frequency, during Q5 it decreased its frequency and was heading towards the location where artifacts were seen during Q2. This may resemble a periodic change of the same artifact caused by the same process. Fig. 10 presents the 20+ artifact.

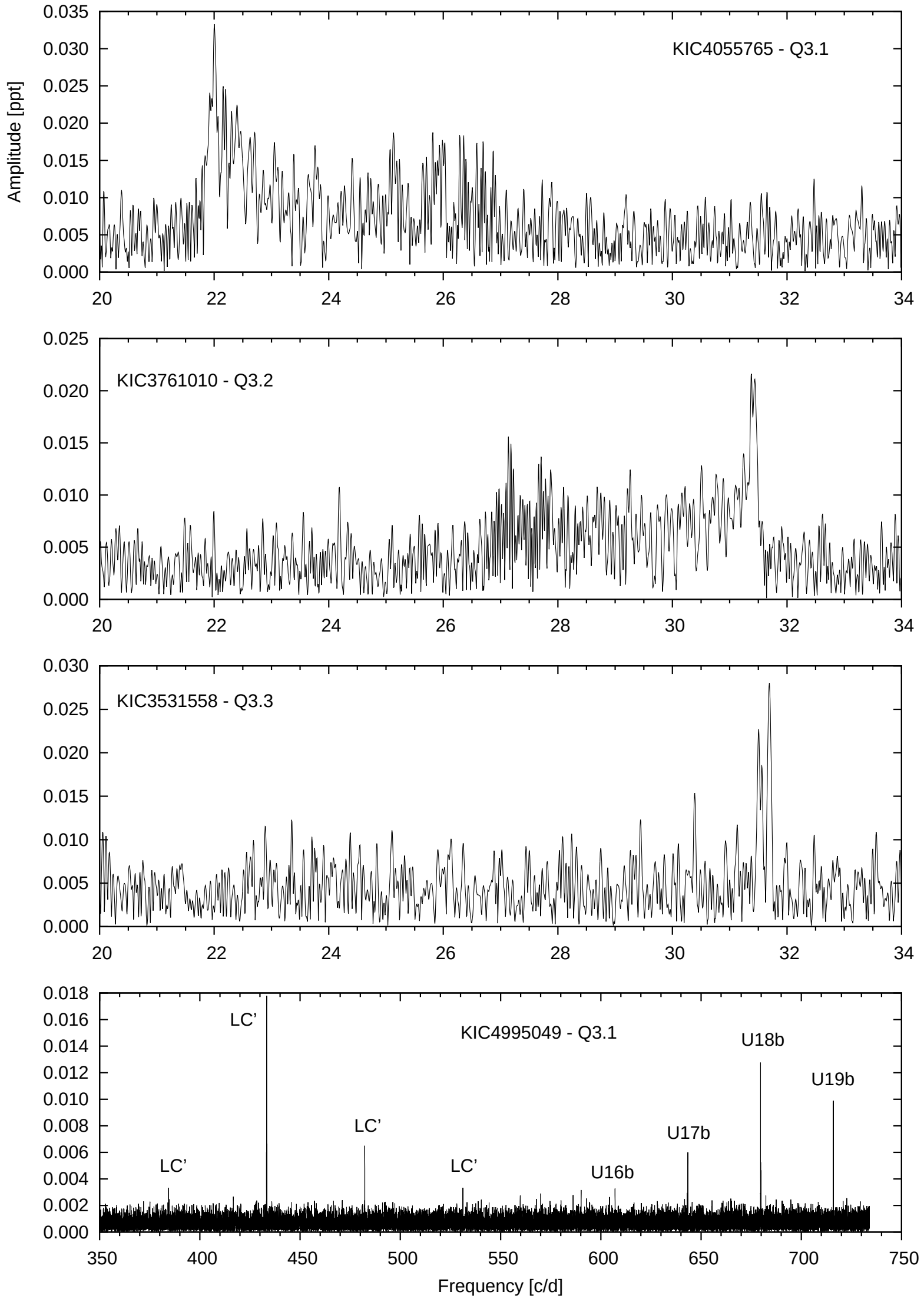

Fig. 8. Close-ups of the amplitude spectra of a few stars calculated from Q3 to show the wide 20+ artifact (*top three panels*) and the LC' comb (*bottom panel*) along with the U sequence.

Q6: We detected the U sequence for most of the stars and the common artifact at 14-22 c/d depending on the month. The frequencies of the wide artifact are smaller by 1 c/d as compared to Q2. We show this artifact in Fig. 11. Q7: This is another quarter where, besides a few cases of the U sequence and the W artifact, a strong, very common wide 20+ artifact appeared (Fig. 12). It changes its shape and frequency from month to month and it becomes clear that the artifacts may monotonically float in frequency over the quarter. The frequency was 23 c/d on average during the first month, then increased to 28 c/d during month 2 and stabilized at 31.5 c/d during the final month. A similar situation happened during previous quarters, however the direction and the frequency range are different.

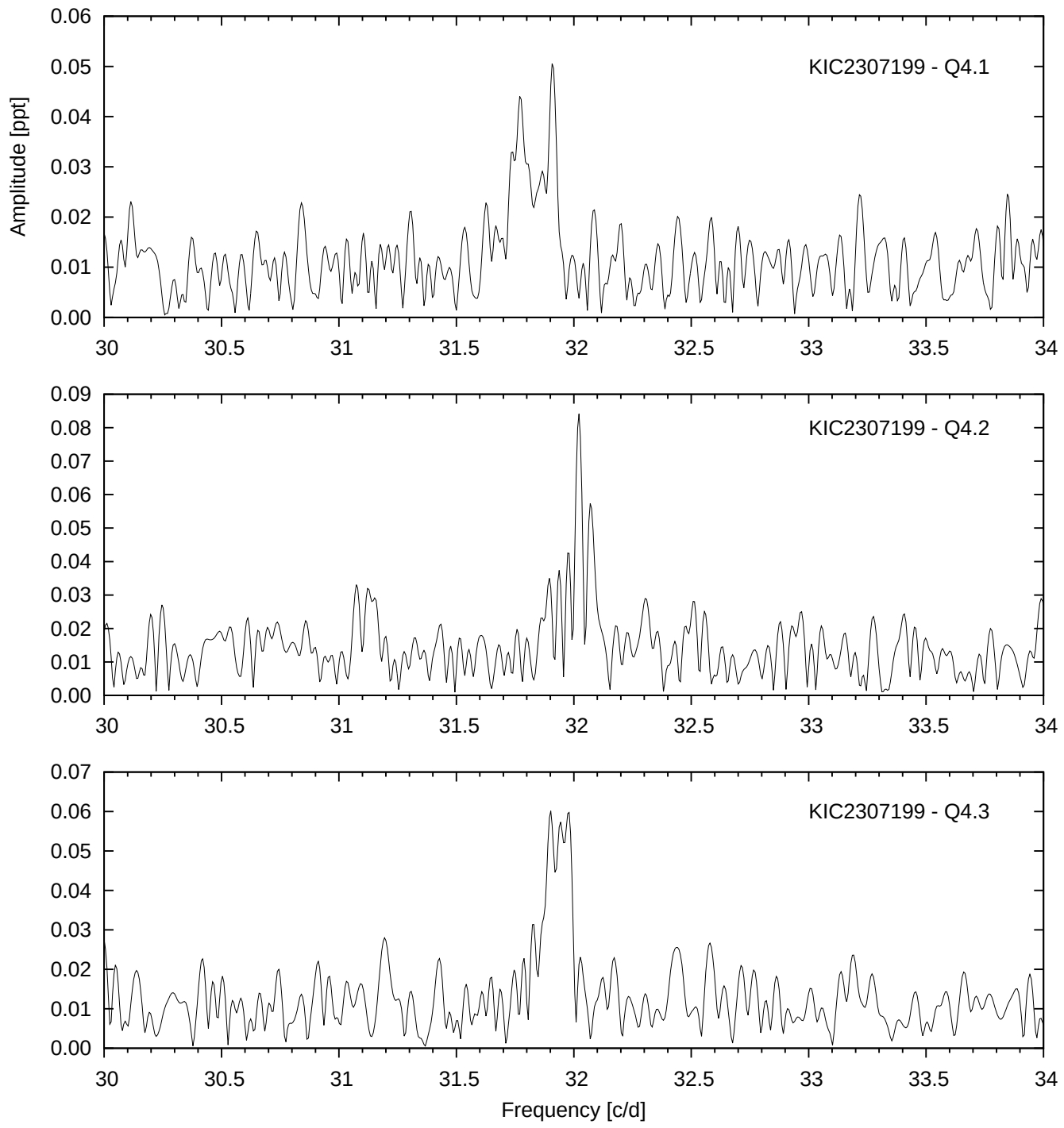

Fig. 9. Close-ups of the amplitude spectrum of KIC 2307199 calculated from Q4 to show the 20+ artifact for separate months.

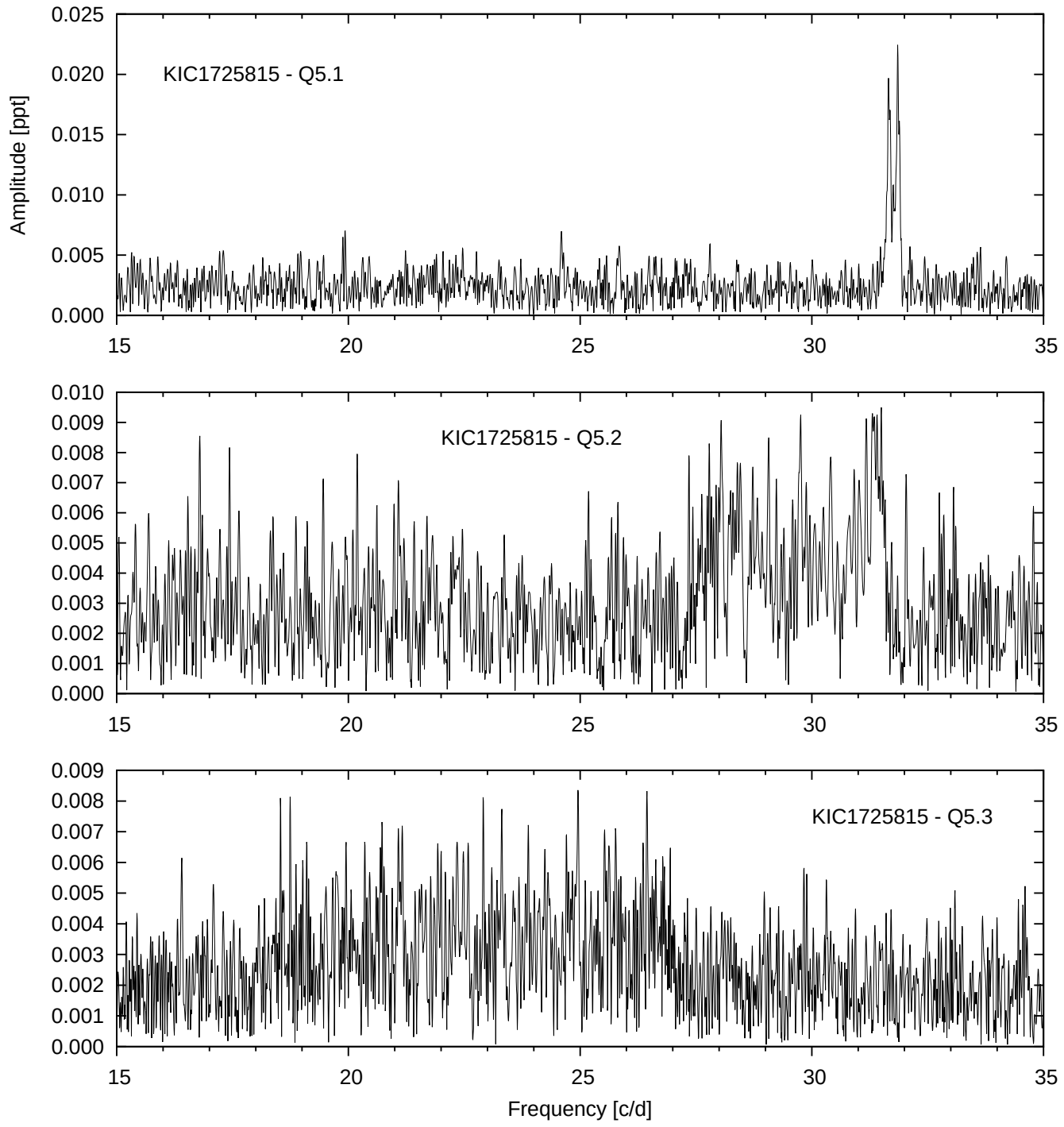

Fig. 10. Close-ups of the amplitude spectrum of KIC 1725815 calculated from Q5 to show the wide 20+ artifact for separate months (*top* to *bottom*).

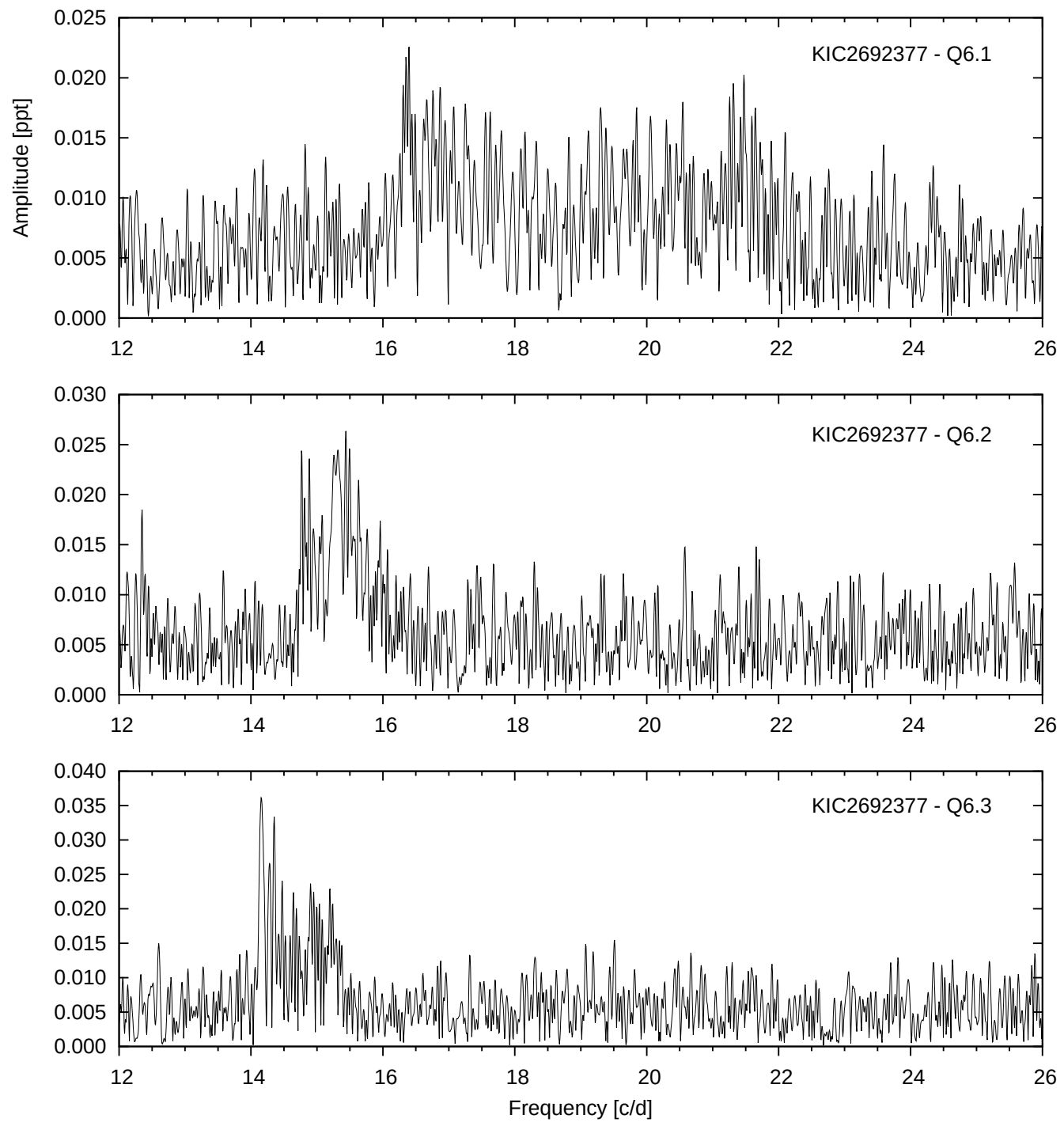

Fig. 11. Close-ups of the amplitude spectrum of KIC 2697388 calculated from Q6 to show the wide 20- artifact for separate months (*top* to *bottom*).

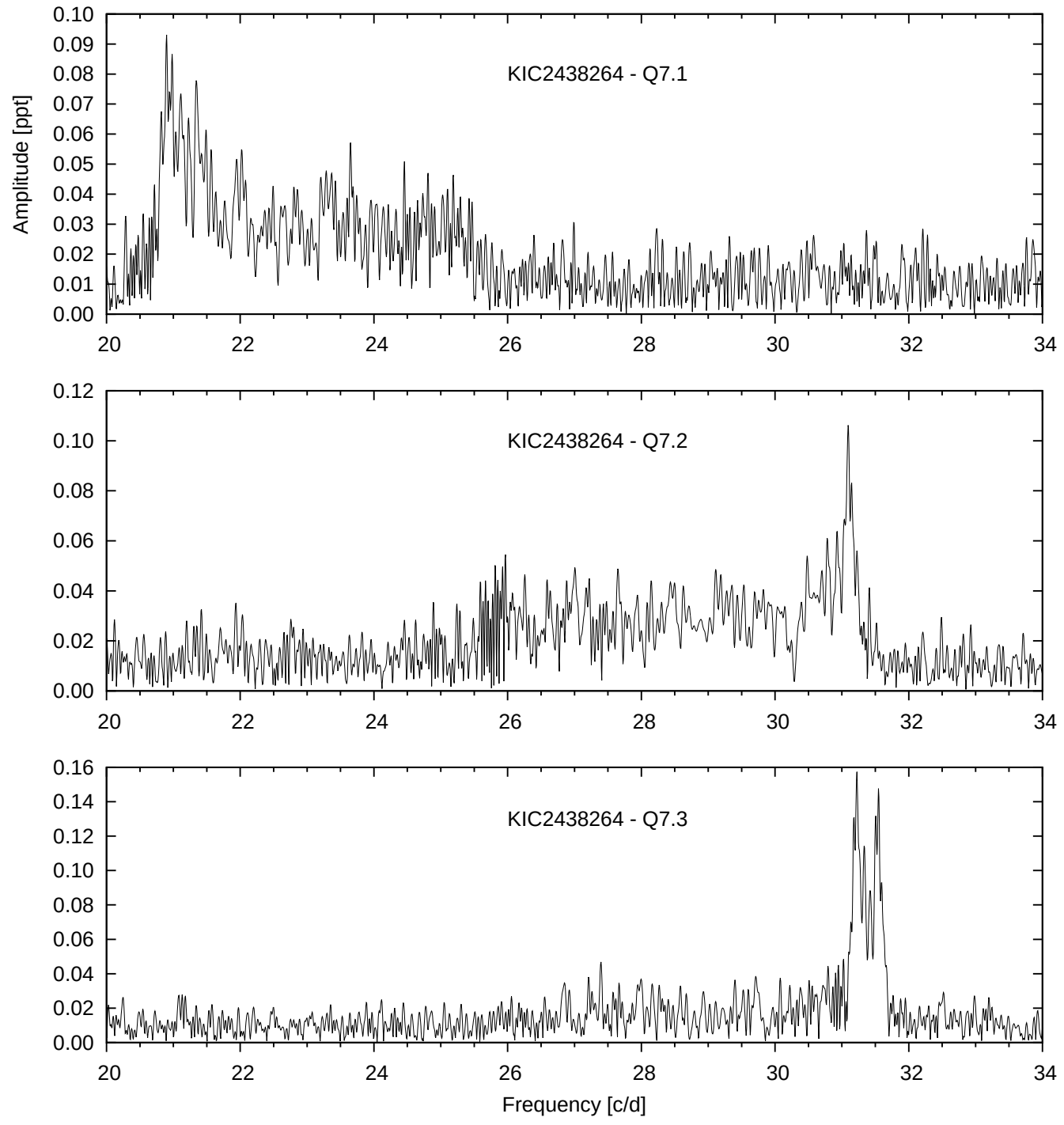

Fig. 12. Close-ups of the amplitude spectrum of KIC 2438264 calculated from Q7 to show the wide 20+ artifact for separate months (*top* to *bottom*).

Having analyzed Q2 up to Q7 we can follow the trend of the wide artifacts' changes, considering them to be the same event (explained in Section 4). During Q2 it decreased from 23 c/d and stabilized at 16 c/d. In Q3 it moved from 23 c/d up and stabilized at 31.7 c/d. During the next quarter, the wide artifact stayed at approximately 32 c/d, decreasing again from 32 c/d down to 19 c/d during Q5. Q6 is a mirror view of Q2 and so is Q7 with respect to Q3. It appeared that there is a four-quarter period of that wide artifact change and this should naturally be associated with a full roll of the spacecraft. The time scales (periodicities) of these artifacts range from 45 min up to 2 hr. We have not found any similar timescales associated with the spacecraft. If the four-quarter period is, in fact, related to the temperature variation (Fig. 10 in Christiansen *et al.* 2013) then we could potentially expect that the short periodicities (between 45 min and 2 hs) may also be caused by temperature instability. Unfortunately the temperature readouts are too sparse making our anticipation impossible to verify.

Q8 and Q9: Following the above conclusion, in Q8 (like Q4) we can expect relatively sharp profiles of the 20+ artifact. In fact, the expected artifact is at the right position and we show it in Fig. 13. Its amplitude varies from star to star, however in the case of the presented KIC 6106120 it has a very large amplitude. In Q9, as expected, we found the same trend of the artifact detected in Q5. We show it in Fig. 14. The 157 c/d artifact is now at 155.6 c/d.

Q10: We found the 20- artifact at about 13 c/d. This confirms the 4 quarters periodic cycle of the artifact variability. Apart from Q2 and Q6 we did not detect any 20- artifact during the first month. During the next two months the artifact's behavior is as expected, although its frequencies are a bit smaller as compared to Q6. They differ by 2 c/d. We have also detected the U sequence which was supposed to be strong and very common during this quarter. We found more peaks of that sequence than in the previous quarters. We show the 20- artifact and the U sequence in Fig. 15.

Q11, Q12 and Q13: These three quarters show the same artifacts as Q7, Q8 and Q9. This confirms our periodicity in the wide 20+ artifact variation on a time scale of four spacecraft rolls. The U sequence remains quiet and was noticeable only in a few cases. We show the shape of the wide artifact in Figs. 16, 17 and 18, respectively for Q11, Q12 and Q13.

Q14: According to the derived periodicities of the artifacts we anticipated the wide 20- and the U sequence to be commonly found during this quarter. Indeed, the U sequence is present while the wide artifact follows its behavior from Q10. The latter means that the wide artifact is absent in the first month of Q14. In addition,

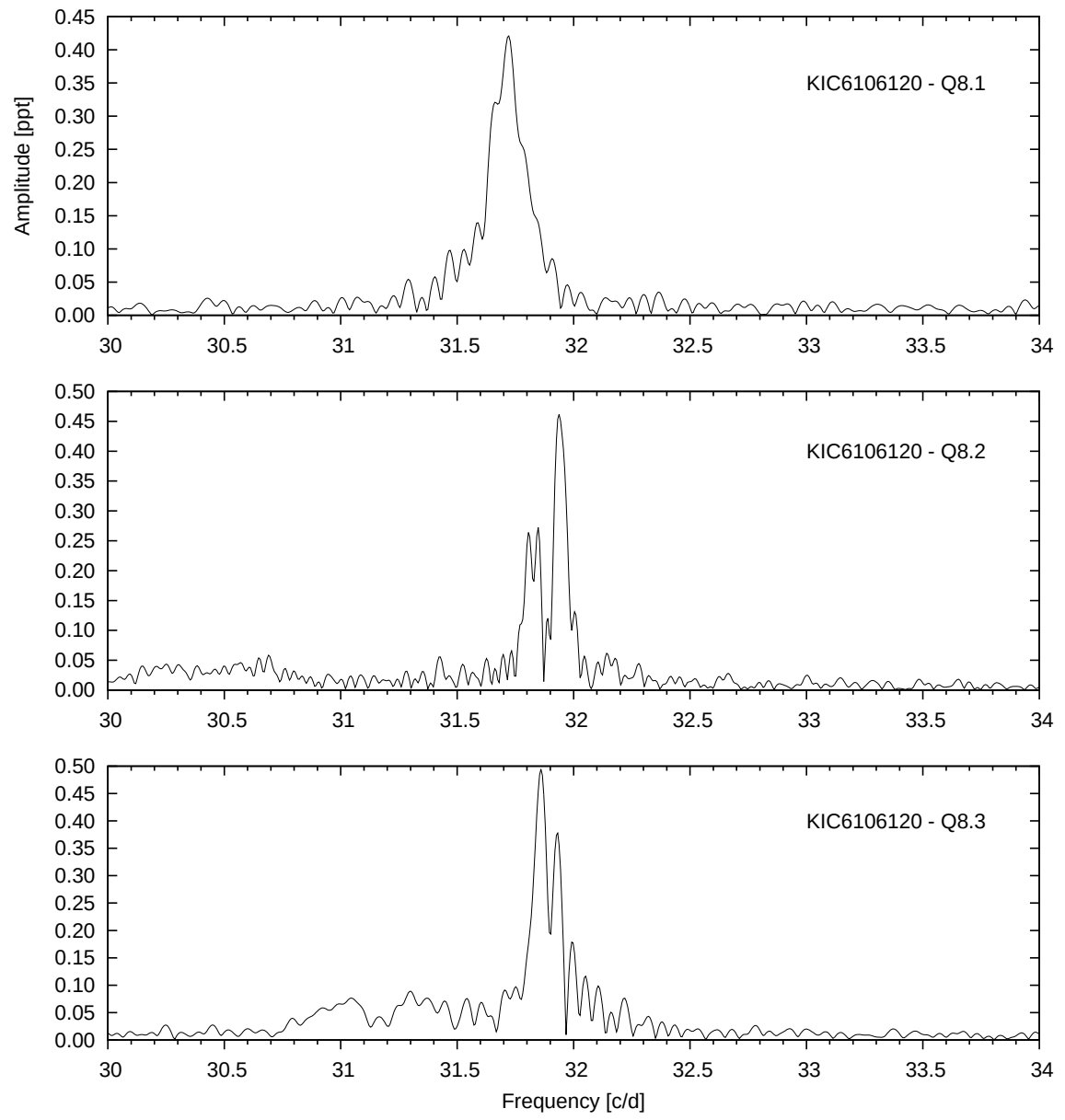

Fig. 13. Close-ups of the amplitude spectrum of KIC 6106120 calculated from Q8 to show the 20+ artifact for separate months (*top* to *bottom*). In case of this star the artifact is very strong in amplitude.

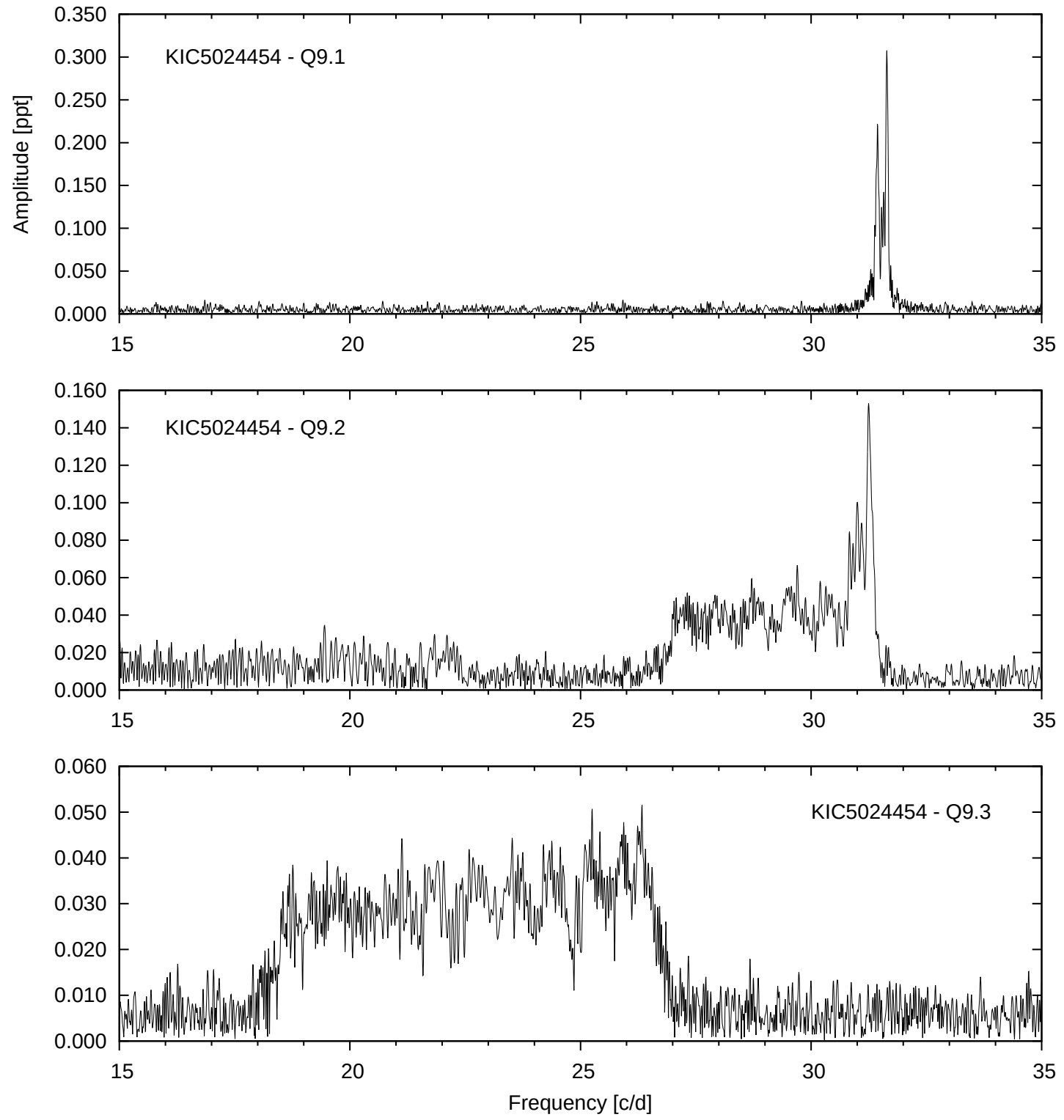

Fig. 14. Close-ups of the amplitude spectrum of KIC 5024454 calculated from Q9 to show the 20+ artifact for separate months (*top* to *bottom*). Similarly to KIC 6106120, the artifact is very strong in amplitude.

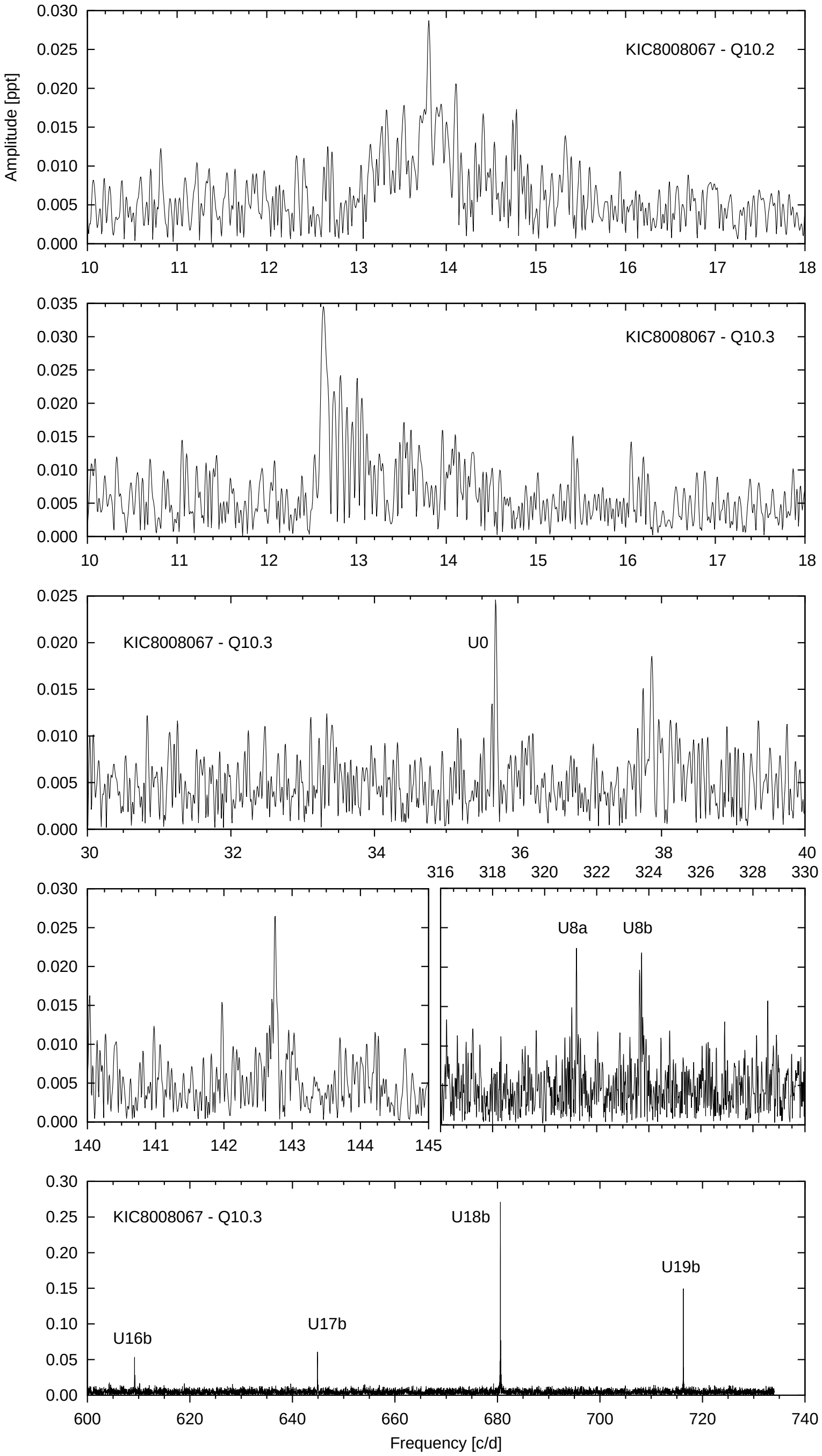

Fig. 15. Close-ups of the amplitude spectrum of KIC 8008067 calculated from Q10 to show the 20+ artifact for separate months (*top two panels*). *Three bottom panels* show the U sequence, particularly the parent peak at 35.7 c/d (*middle panel*) as well as co-existence of two peaks from two separate U sequences (a and b) at around 322 c/d.

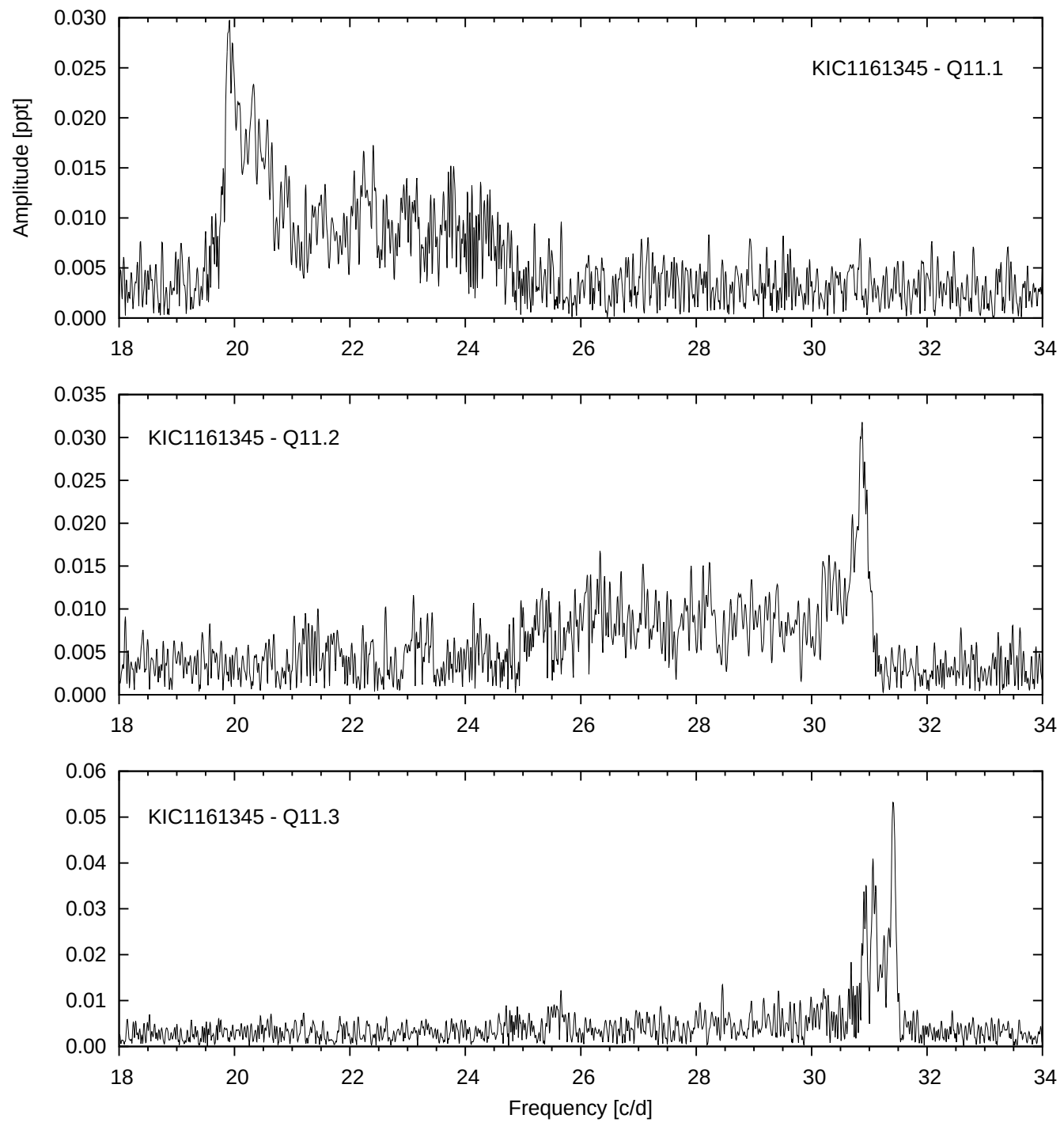

Fig. 16. Close-ups of the amplitude spectrum of KIC 1161345 calculated from Q11 to show the 20+ artifact for separate months.

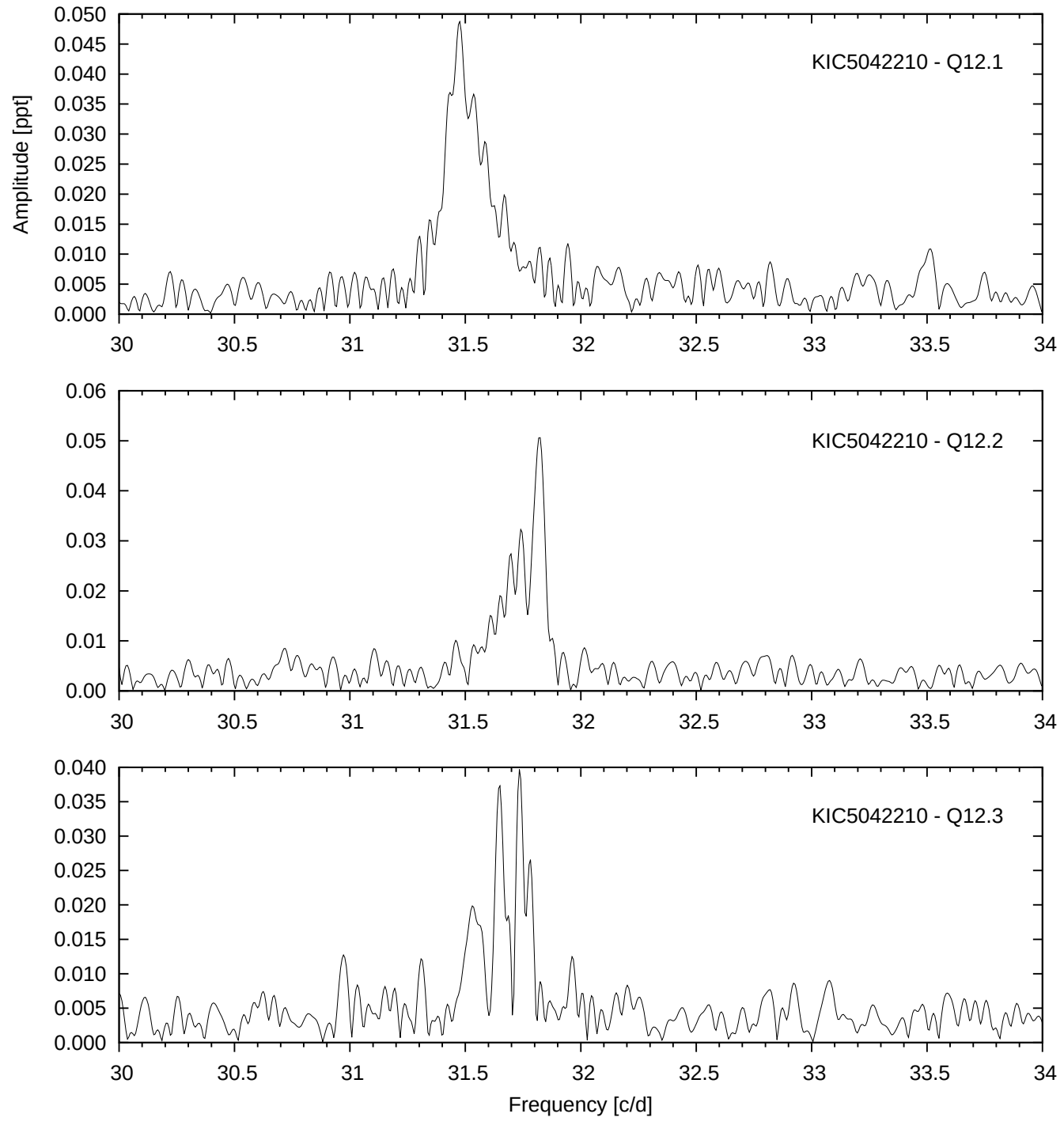

Fig. 17. Close-ups of the amplitude spectrum of KIC 5042210 calculated from Q12 to show the 20+ artifact for separate months.

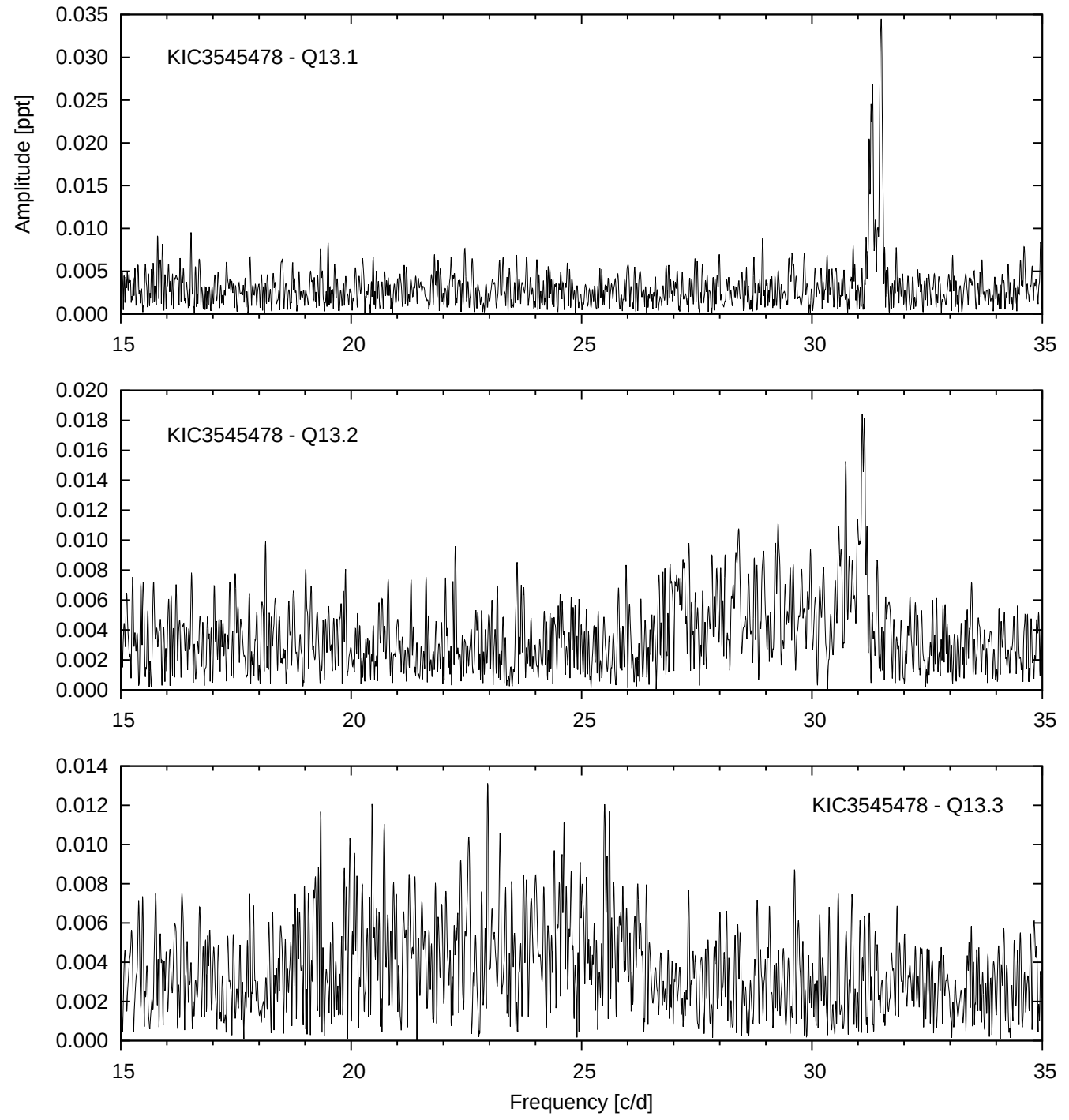

Fig. 18. Close-ups of the amplitude spectrum of KIC 3545478 calculated from Q13 to show the 20+ artifact for separate months.

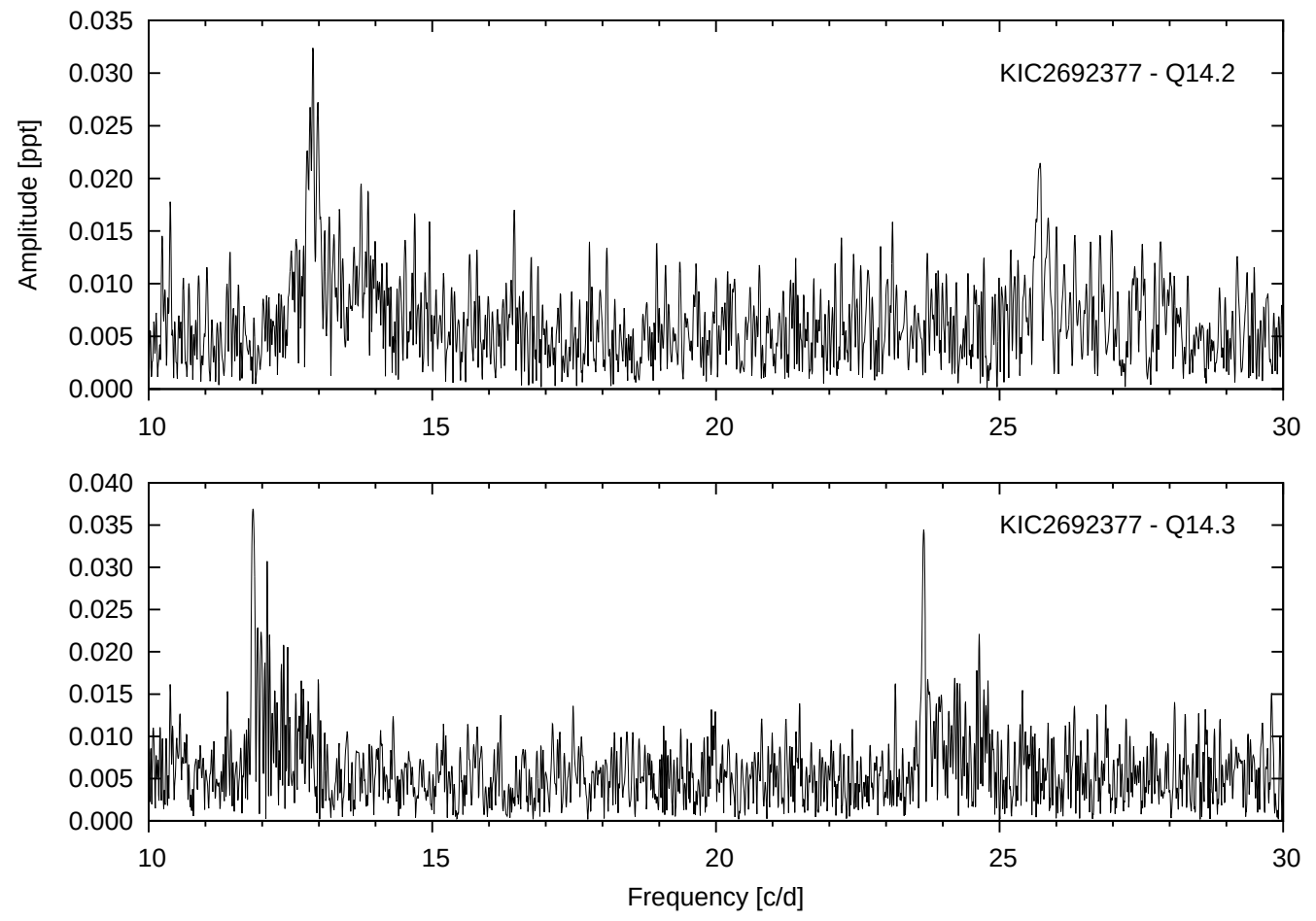

Fig. 19. Close-ups of the amplitude spectrum of KIC 2692377 calculated from Q14 to show the 20- artifact for separate months. It was not detected during the first month.

the frequencies in the consecutive months are smaller by 1 c/d as compared to Q10. Since Q2 this artifact has changed its frequencies by 4 c/d in the second and third month of a given quarter, while disappeared during the first one. Fig. 19 presents the wide artifact detected during the second and the third months.

## 4. Is the Wide Artifact Wide?

A periodic variation which is fixed in frequency/phase is represented in an amplitude spectrum by a sharp peak at a given frequency. This is not true if the above parameters of that variation vary. In such case, the peak will become spread over a range of frequencies, depending on the level of frequency/phase change. The wide artifact we detected in Kepler data may be explained in two ways. Either its frequency/phase is changing so severely contributing to a dramatic smudge in the amplitude spectrum or it is a set of simultaneous close, sharp (stable in frequency) peaks which overlap creating a wide hump instead. One way to sort this out is to split data into shorter (than a month) chunks and verify how the artifact behaves on a shorter time scale. In Fig. 20 we show the time frequency diagram spread over all data accessible to us. The diagram was derived in the same manner as the one presented in Fig. 2. It clearly shows that the two wide artifacts are continuously changing their frequency contributing to the smudges when the monthly data cover large frequency evolution and to the rather sharp peaks when the artifacts in that diagram are mostly represented by a vertical line.

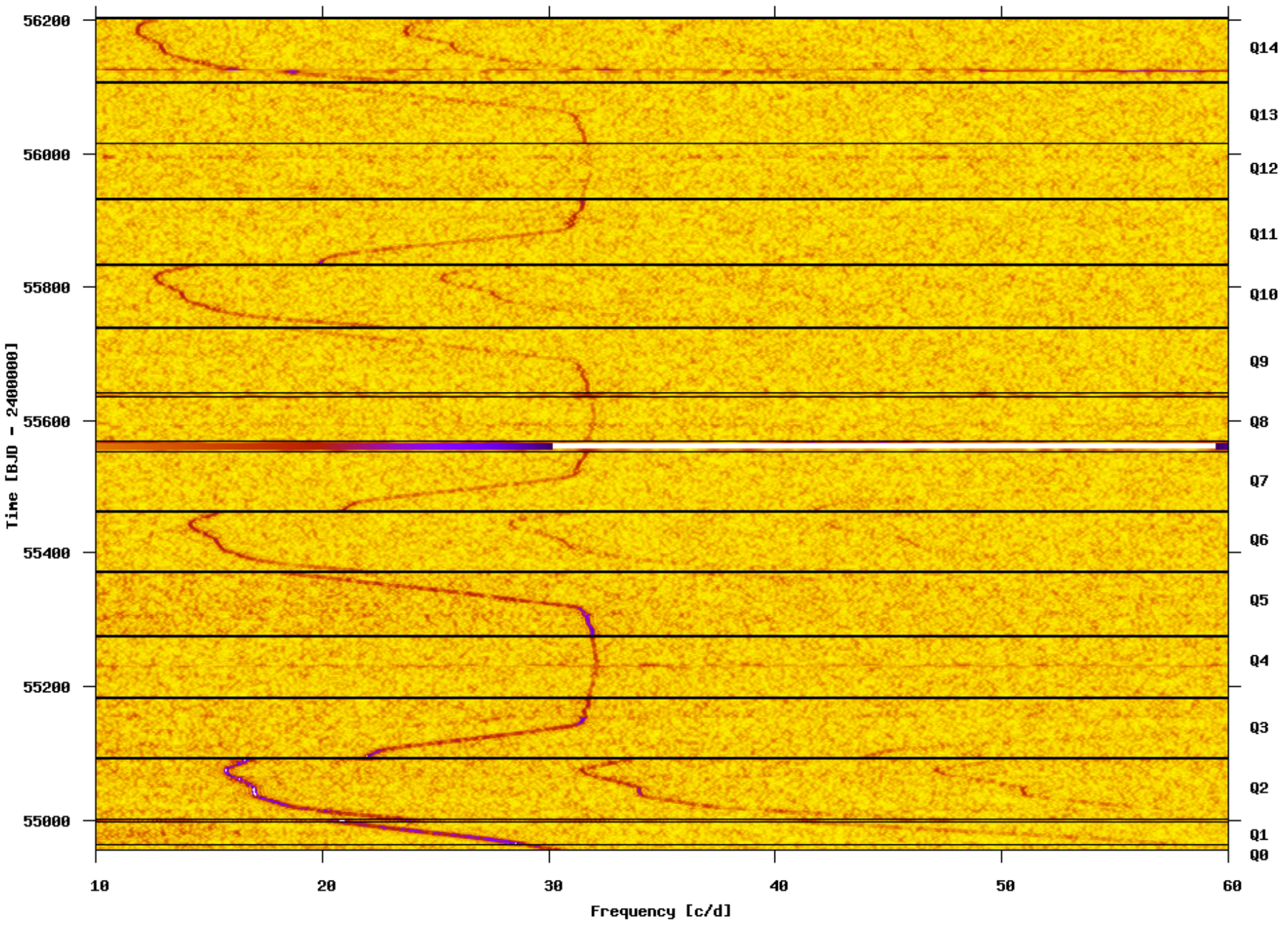

Fig. 20. Time frequency diagram showing the evolution of the artifacts between 10 and 60 c/d.

From the diagram we can see that the 20+ artifact is stable between quarters. It always starts and ends at the same frequencies, moreover its turning side is also stable. Instead, the 20- artifact may still start at the same frequency from quarter to quarter, however its turning point is clearly moving towards lower frequencies indicating that the time scale is increasing. The harmonics appear to be stronger during the beginning of the mission. We detected at least two harmonics of the 20- artifact and the first one of 20+ artifact during Q2.

These two wide artifacts could have the same origin. In Fig. 20 we see that they do not close the loop but slightly overlap in frequency when 20+ is swapping into 20- and there is a discontinuity when 20- ends. These discontinuities always occur during quarterly rolls. The overlapping part is somehow comparable in frequency to the gap between 20- and 20+. In addition the part above 20 c/d is stable between quarters while the part below 20 c/d is not. On the other hand, both artifacts complete a cycle suggestive of the same source with a possible instability during some quarters (Q2, Q6, Q10, Q14, ...).

## 5. Summary

Our analysis clearly revealed that besides LC artifacts there is also the U sequence, a few peaks separated by LC readout but offset by almost 7 c/d from the regular LC comb, the W artifact as well as the wide artifacts around 20 c/d. We did not notice any correlation between frequencies of the U sequence and a specific month/quarter, except that they vary. However, the U sequence becomes very common among the stars every four rolls. Basically, there are three quiet quarters and during the fourth one, the U sequence acts like a Jack in the box. We did notice that the U sequence becomes very strong when the light curve looks like the one in Fig. 6. The folded outline pattern shows 3 day periodicity and it may suggest that the U sequence is related to the reaction wheels desaturation or temperature instability.

The wide artifacts experience very complex evolutions during one year. In Fig. 20 we present their variations which is self-explanatory as concerns their wideness and spikiness when analyzed on a monthly basis. We concluded that the artifacts around 20 c/d are periodically variable on a time scale of one year or four spacecraft rolls. This could suggest that the artifact is somehow associated with the spatial orientation in space, likely with respect to the Sun which shines on the spacecraft at different angles causing a differential heating of the electronics onboard. This is, however, a slow process and likely not the reason for the artifacts' evolutions. With every roll the stars fall onto different CCDs (except module 13), however it is not essential since we noticed the same shape of the artifacts within a given month for most of the stars even located on different CCDs (Fig. 21).

Searching for correlation between time scales of the artifacts and possible events on board we have not spotted any correlation between artifacts and a specific CCD or pipeline reduction (either SAP or PDC). Not all the stars are affected by artifacts. Those with no artifacts are just a few, though, many stars do not show strong effects by any of the reported artifacts. This may be caused by the fact that only certain pixels are affected by artifacts and when these pixels are included in the optimal apertures the artifacts will show up in the amplitude spectra. This brings us to the conclusion that brighter stars (more pixels) should show stronger artifacts. One must be aware that this does not have to be a rule since it may always happen that none of those infected pixels are in the aperture. In addition, the pixels do not bear artifact's signatures all the time.

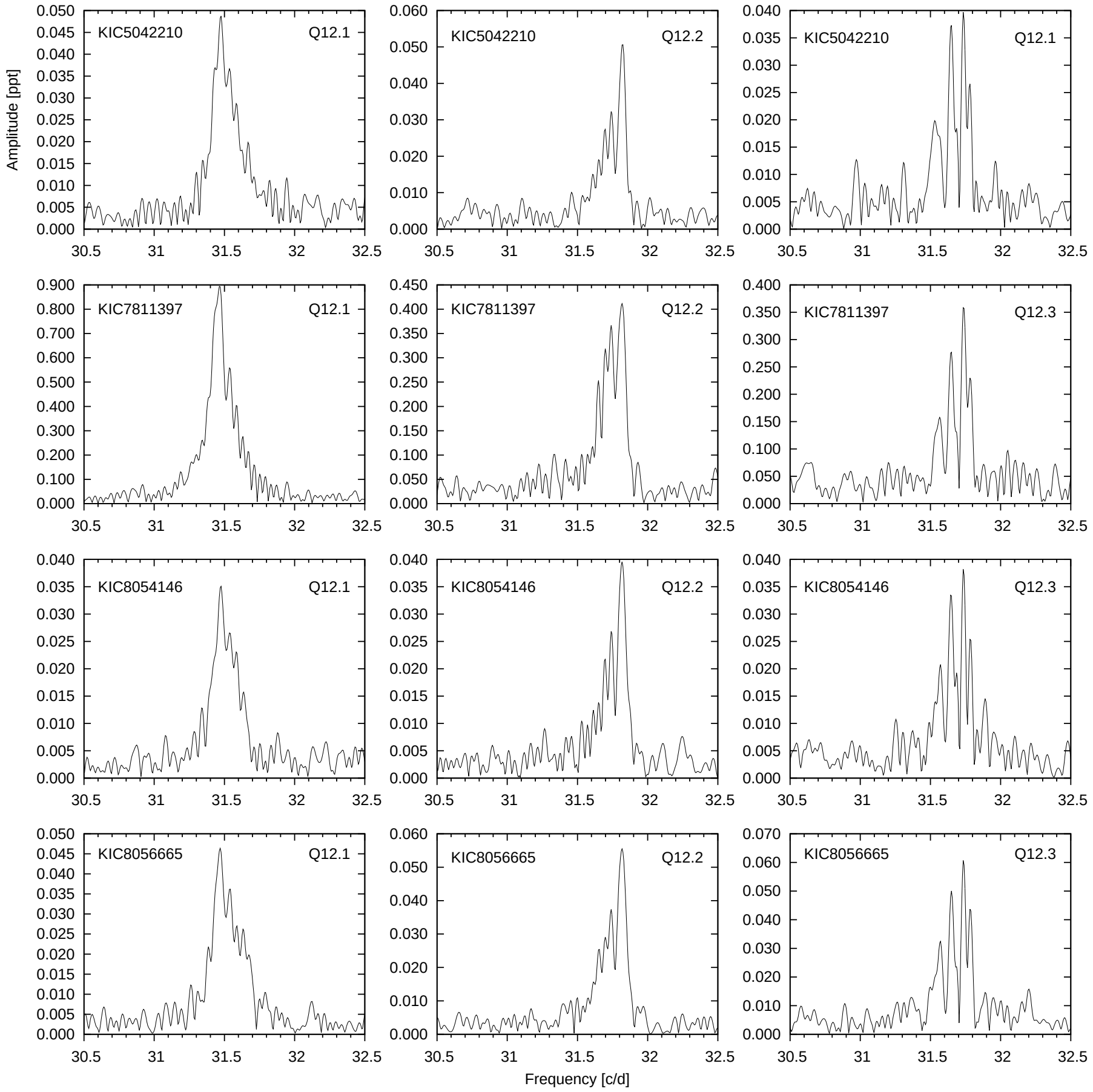

Fig. 21. Mosaic of panels showing that the shape of the artifacts remain similar among stars during a specific month.

Unfortunately, we cannot provide definite sources of the artifacts we described in this paper. However, the goal was two-fold. First, to analyze SC data to pull out the artifacts and, if possible, provide their sources to mitigate them in the near future. Second, to illustrate the artifacts to help other investigators to eliminate those "fake" periodicities from their solutions and by that to avoid a misinterpretation of the data. We found many targets with the artifacts hidden among the real signal (two shown in Fig. 1). Data Release Notes attached to quarterly data releases already provide the list of the artifacts, however, stars may show a real signal at or close to the artifacts and having only the frequencies could be confusing in eliminating potential artifacts. Once plotted and knowing that their shapes remain similar among the stars it is easier to exclude "fake" signal from the solution. To properly distinguish between the artifacts and real signal, one has to analyze data on a monthly basis and check if an artifact characteristic for a given month is present in the data. If not, even though there is a peak in an amplitude spectrum calculated from all the available data at an artifact's frequency, the peak should be considered as real.

The analyses of different types of pulsating stars may be affected by different artifacts. Among compact pulsators, white dwarfs are affected by U, LC and LC' sequences while subdwarf B stars by all the artifacts since most of the latter pulsators are hybrids pulsating at both low and high frequencies (from a few c/d up to the Nyquist frequency of SC data). Main sequence stars, including solar-like oscillators (along with red giants), $\beta$ Cep, $\delta$ Sct and SX Phe, will be affected by wide artifacts and their harmonics. The exceptions here are rapidly oscillating Ap stars which pulsate at minutes pace and will only be affected by U, LC, LC' sequences. Slowly pulsating B stars, $\gamma$ Dor and Cepheids may only be affected by 0.33 c/d artifact. RR Lyr star may be affected by 0.33 c/d and wide artifacts.

Table 3

Appearances of the artifacts during the quarters of data collection

| ID | 0 | 1 | 2 | 3 | 4 | 5 | 6 | 7 | 8 | 9 | 10 | 11 | 12 | 13 | 14 |
|---|---|---|---|---|---|---|---|---|---|---|---|---|---|---|---|
| 20- | - | - | + | - | - | - | + | - | - | - | + | - | - | - | + |
| 20+ | + | + | - | + | + | + | - | + | + | + | - | + | + | + | - |
| U | + | + | +++ | + | + | + | +++ | + | + | + | +++ | + | + | + | +++ |

"+++" for U sequence means that they become very common and strong every four quarters.

**Acknowledgements.** This project was supported by the Polish National Science Centre under project No. UMO-2011/03/D/ST9/01914 and US NSF grant AST-1009436. The author gratefully thanks to Fergal Mullally, Ron Gilliland and Amanda Winans for reading the manuscript and their generous suggestions. The comments and suggestions from the anonymous referee are also appreciated. Funding for this Discovery mission is provided by NASA's Science Mission Directorate. The authors gratefully acknowledge the entire Kepler team, whose efforts have made these results possible.